\documentclass[epsfig]{article}
\usepackage{amsfonts}
\usepackage{graphicx}
\usepackage{amsmath}

\input epsf.sty
\newtheorem{theorem}{Theorem}
\newtheorem{acknowledgement}[theorem]{Acknowledgement}

\input{tcilatex}

\begin{document}

\title{\rightline{\mbox {\normalsize
{Lab/UFR-HEP0508/GNPHE/0508}}} \textbf{Calogero model with Yukawa like
interaction}}
\author{Mohammed Kessabi$^{1,2}$, El Hassan Saidi$^{1,2,3}$\thanks{%
h-saidi@fsr.ac.ma} , Hanane Sebbata$^{1,2}$ \\
{\small \textit{1.}} {\small \textit{Lab/UFR-Physique des Hautes Energies,
Facult\'{e} des Sciences de Rabat, Morocco.}}\\
{\small \textit{2. Groupement National de Physique des Hautes Energies,
GNPHE; }}\\
{\small \textit{Siege focal, Lab/UFR-HEP, Rabat, Morocco.}}\\
{\small \textit{3. VACBT, Virtual African Centre for Basic Science and
Technology, }}\\
{\small \textit{Focal point Lab/UFR-PHE, Fac Sciences, Rabat, Morocco.}}}
\maketitle

\begin{abstract}
We study an extension of one dimensional Calogero model involving strongly
coupled and electrically charged particles. Besides Calogero term $\frac{g}{%
2x^{2}}$, there is an extra factor described by a Yukawa like coupling
modeling short distance interactions.\ Mimicking Calogero analysis and using
developments in formal series of the wave function $\Psi \left( x\right) $
factorised as $x^{\epsilon }\Phi \left( x\right) $ with $\epsilon \left(
\epsilon -1\right) =g$, we develop a technique to approach the spectrum of
the generalized system and show that information on full spectrum is
captured by $\Phi \left( x\right) $ and $\Phi ^{\prime \prime }\left(
x\right) $ at the singular point $x=0$ of the potential. Convergence of $%
\int dx\left\vert \Psi \left( x\right) \right\vert ^{2}$ requires $\epsilon
>-\frac{1}{2}$ and is shown to be sensitive to the zero mode of $\Phi \left(
x\right) $ at $x=0$.

\end{abstract}

\tableofcontents

\newpage

\section{Introduction}

\qquad The study of integrability of hamiltonian systems has been one of the
basic tasks during decades as it gives precise informations on classical and
quantum dynamics as well as spectrum of the hamiltonians $\cite{1,2}$. This
analysis has lead to the discovery of several features regarding quantum
spectrum of integrable hamiltonian systems $\cite{3}$-$\cite{10}$. The
familiar examples are given by the remarkable class of integrable low
dimensional field models based on both finite dimensional and affine Lie
algebras. The study of integrable hamiltonian systems has also shown that
there is only a few number of real physical systems that are completely
solvable. The usual harmonic oscillator and one dimensional Calogero model
as well as extensions $\cite{11}$-$\cite{14}$ are certainly some of these
examples. Despite the variety of methods used to study integrability, one
may, roughly speaking, say that the main difficulty in getting exact
solutions comes essentially from the fact that the underlying differential
equations are non linear coupled relations difficult to solve. Lessons
learnt from the study of hamiltonian systems and infinite dimensional
symmetries of integrable models let understand that hamiltonian systems may
in general be classified into three basic sets: (\textbf{i}) integrable
hamiltonian systems where the spectrum of the hamiltonian is exactly
determined as in case of Calogero model $\cite{3}$, (\textbf{ii})
quasi-integrable systems where, though not completely determined, the
hamiltonian spectrum, obtained by perturbation of exact models, is under
control and (\textbf{iii}) remaining others with unknown spectrum.

\qquad Motivated by remarkable properties of integrable one dimensional
Calogero type interaction $\cite{4}$, we study in this paper the case where
Calogero particles are moreover electrically charged and strongly coupled.
Coulomb interaction $\frac{1}{\left\vert x_{i}-x_{j}\right\vert }$ and
strong coupling between pairs of particles $\left\{ x_{i},x_{j}\right\} $
are implemented by adjunction to the usual Calogero hamiltonian an extra
short distance interaction described by a Yukawa like coupling $\exp \left(
-a\left\vert x_{i}-x_{j}\right\vert \right) /\left\vert
x_{i}-x_{j}\right\vert $ where $a$ is the inverse of Debye length $\lambda $%
. Among our results, we show that for the case of two particles, the
discrete spectrum $\left\{ E_{n},\Psi _{n}\left( x\right) =x^{\epsilon }\Phi
_{n}\left( x\right) \right\} $ of this generalized Calogero system with $%
\epsilon >-\frac{1}{2}$ is completely determined by the knowledge of the
wave factor $\Phi _{n}\left( x\right) $ and $\Phi _{n}^{\prime \prime
}\left( x\right) $ at the singular point $x=0$ of the potential. The range $%
\epsilon >-\frac{1}{2}$ is required by the convergence of $\int dx\left\vert
\Psi _{n}\left( x\right) \right\vert ^{2}$ and is shown to be sensitive to
the zero mode of $\Phi _{n}\left( x\right) $ at $x=0$.

\qquad The organisation of this paper is as follows: In section 2, we review
briefly quantum Calogero model by solving directly the Calogero wave
equation for the case of two particles; the case of $N$ particles is
straightforward and follows the same lines. Comments regarding this solution
are also given; they are used in the interpretation of our generalization.
In section 3, we study the extension of Calogero model involving charged and
strongly correlated particles described by a Yukawa like interaction.
Motivated by similarities with Calogero analysis and inspired from moment
technique, we develop a method to approach the discrete spectrum of the
generalized Calogero system. In section 4, we give our conclusion and make a
discussion regarding some specific limits.

\section{$1D$ Calogero model: brief review}

\qquad One dimensional Calogero model describing the quantum dynamics of two
interacting particles, of local coordinates $x_{1}$ and $x_{2}$ and relative
position $x=\left( x_{2}-x_{1}\right) /\sqrt{2}$, is governed by the
following eigenvalue wave equation $\mathcal{H}\Psi \left( x\right) =E\Psi
\left( x\right) $,%
\begin{equation}
\left( -\frac{1}{2}\frac{d^{2}}{dx^{2}}+\frac{1}{2}\omega ^{2}x^{2}+\frac{g}{%
2x^{2}}\right) \Psi \left( x\right) =E\Psi \left( x\right) ,  \label{ca}
\end{equation}%
where to fix the idea we take $x>0$; i.e $x_{2}>x_{1}$. In this relation, $%
\omega $\ is the usual harmonic oscillator frequency and $g$ is the Calogero
coupling parameter. To get the discrete spectrum $\left\{ E_{n},\Psi
_{n}\left( x\right) \right\} $ of this differential equation, it is
interesting to use the following direct method which is also valid for the
case of a system of $N$ particles. This method relies on solving directly eq(%
\ref{ca}) and involves three steps each one associated with the
factorisation of the field variable $\Psi \left( x\right) $ as $%
x^{\varepsilon }\Phi \left( x\right) =x^{\varepsilon }P\left( x\right) \exp
\left( -\frac{\omega }{2}x^{2}\right) $ for some modulus $\epsilon $. First,
set the change $\Psi \left( x\right) =x^{\varepsilon }\Phi \left( x\right) $%
, which for $\epsilon >0$, reflects just the fact that two classical
particles $\left\{ x_{1},x_{2}\right\} $ cannot live together at $%
x_{1}=x_{2} $; the wave function should then vanish; i.e $\Psi \left(
0\right) =0$; but $\Phi \left( 0\right) \neq 0$. Note that the choice $%
\epsilon >0$ is not a necessary condition, since quantum mechanically we
should rather have,%
\begin{equation}
\int_{0}^{\infty }dx\left\vert \Psi \left( x\right) \right\vert ^{2}<\infty .
\label{con}
\end{equation}%
As far convergence of this integral is concerned, one should not that there
are two sources of possible divergences, one for $x\rightarrow 0$ and the
other for $x\rightarrow \infty $. Decomposing eq(\ref{con}) as $%
\int_{0}^{y}dx\left\vert \Psi \left( x\right) \right\vert
^{2}+\int_{y}^{z}dx\left\vert \Psi \left( x\right) \right\vert
^{2}+\int_{z}^{\infty }dx\left\vert \Psi \left( x\right) \right\vert ^{2}$
for some small positive number $y$ and a large positive $z$; then using the
fact that $\int_{z}^{\infty }dx\left\vert \Psi \left( x\right) \right\vert
^{2}$ is finite\footnote{%
For large $x$, say $z\leq x<\infty $, Yukawa like potential may be neglected
with respect to Calogero interaction. So $\int_{z}^{\infty }dx\left\vert
\Psi \left( x\right) \right\vert ^{2}\rightarrow \int_{z}^{\infty
}dx\left\vert \Psi ^{\left( cal\right) }\left( x\right) \right\vert ^{2}$
which is known to be finite.}, we will prove later that in the region $0\leq
x\leq y$, convergence requires that $\frac{y^{2\varepsilon +1}}{2\varepsilon
+1}\left\vert \Phi \left( 0\right) \right\vert ^{2}$ should be finite. By
substitution of the change $\Psi \left( x\right) =x^{\varepsilon }\Phi
\left( x\right) $ in eq(\ref{ca}), we get the following constraint relation
on $\Phi $,
\begin{equation}
\Phi ^{\prime \prime }\left( x\right) +\frac{2\varepsilon }{x}\Phi ^{\prime
}\left( x\right) +\left( 2E-\omega ^{2}x^{2}\right) \Phi \left( x\right) =0,
\label{no}
\end{equation}%
which, for a matter of discussion of the solution, we prefer to rewrite it
as follows,%
\begin{equation}
\Phi ^{\prime \prime }\left( x\right) +\frac{2\varepsilon }{x}\Phi ^{\prime
}\left( x\right) +2\left( E-U\right) \Phi \left( x\right) =0,  \label{na}
\end{equation}%
with $U\left( x\right) =\omega ^{2}x^{2}/2$. Note that the factorisation $%
\Psi \left( x\right) =x^{\varepsilon }\Phi \left( x\right) $ has been also
introduced to fix the power $\epsilon $ in term of Calogero coupling
constant $g$ as $g=\varepsilon \left( \varepsilon -1\right) $. This relation
follows from the requirement that the change $\Psi =x^{\varepsilon }\Phi $
has to kill the problematic term $\frac{g}{2x^{2}}$ which is one of the two
sources of difficulty (non linearity) in the search for exact solution. The
other difficulty is treated in the second step which consist to make the
local scaling change,%
\begin{equation}
\Phi \left( x\right) =P\left( x\right) \exp \left( -\frac{\omega }{2}%
x^{2}\right) ,  \label{ga}
\end{equation}%
which for matter of generality, we rewrite it also as $\Phi \left( x\right)
=P\left( x\right) \exp \left( -f\left( x\right) \right) $ with $f\left(
x\right) =\frac{\omega }{2}x^{2}=\frac{U}{\omega }$. Using this change, we
can bring eq(\ref{no},\ref{na}) to the form,%
\begin{equation}
P^{\prime \prime }\left( x\right) +\left( \frac{2\varepsilon }{x}-2f^{\prime
}\right) P^{\prime }\left( x\right) +\left( 2E-2U-2\frac{\varepsilon }{x}%
f^{\prime }-f^{\prime \prime }+f^{\prime 2}\right) P\left( x\right) =0
\label{ee}
\end{equation}%
By substituting $2U=f^{\prime 2}=\omega ^{2}x^{2}$, we discover the
following familiar differential equation,%
\begin{equation}
P^{\prime \prime }\left( x\right) +\left( \frac{2\varepsilon }{x}-2\omega
x\right) P^{\prime }\left( x\right) +\left[ 2E-\omega \left( 2\varepsilon
+1\right) \right] P\left( x\right) =0,  \label{la}
\end{equation}%
where now the coefficient $\left[ 2E-\omega \left( 2\varepsilon +1\right) %
\right] $ in front of $P\left( x\right) $ is constant. This is a typical
equation $\cite{15}$ whose solution is given by a special hypergeometric
function which we describe in the next step. Setting $u=\omega x^{2}$ and $%
P\left( x\right) =L\left( u\right) $ and putting back into above equation,
we get the following familiar differential relation,%
\begin{equation}
uL^{\prime \prime }\left( u\right) +\left( \frac{2\varepsilon +1}{2}%
-u\right) L^{\prime }\left( u\right) +\left( \frac{E}{2\omega }-\frac{%
2\varepsilon +1}{4}\right) L\left( u\right) =0.  \label{lb}
\end{equation}%
This differential equation has a known exact solution given by Laguerre
polynomials $L_{n}^{\left( \varepsilon -\frac{1}{2}\right) }\left( u\right) $
provided that the factor $\left( \frac{E}{2\omega }-\frac{2\varepsilon +1}{4}%
\right) $ is quantized, i.e $\left( \frac{E}{2\omega }-\frac{2\varepsilon +1%
}{4}\right) =n\in \mathbb{N}$,
\begin{equation}
L_{n}^{\alpha }\left( u\right) =\frac{1}{n!}u^{-\alpha }e^{u}\frac{d^{n}}{%
du^{n}}\left[ u^{\alpha +n}e^{-u}\right] ,  \label{lc}
\end{equation}%
where the upper index $\alpha $ is a convention notation; it should not be
confused with a power. Therefore the discrete spectrum $\left\{ E_{n},\Psi
_{n}\left( x\right) \right\} $ of quantum Calogero two particle system reads
as follows:%
\begin{eqnarray}
E_{n}^{cal} &=&\left( 2n+\epsilon +\frac{1}{2}\right) \omega ,\qquad n\in
\mathbb{N},  \notag \\
\Psi _{n}\left( x\right) &=&x^{\epsilon }\exp \left( -\frac{\omega }{2}%
x^{2}\right) L_{n}^{\left( \varepsilon -\frac{1}{2}\right) }\left( \omega
x^{2}\right) .  \label{ma}
\end{eqnarray}%
The ground state is given by $\Psi _{0}\left( x\right) =x^{\epsilon }\exp
\left( -\frac{\omega }{2}x^{2}\right) $ and $E_{0}^{cal}=\left( \epsilon +%
\frac{1}{2}\right) \omega $; for $\epsilon =0$ ($g=0$); one recovers the
usual spectrum of the ground state for harmonic oscillator.

\qquad For later use, it is also interesting to note the three following: (%
\textbf{a}) First remark that the above solution may be also rewritten in a
positive modes series as follows,%
\begin{equation}
\mathcal{\Phi }_{n}^{\left( cal\right) }(x;\epsilon ,\omega
)=\sum_{k=0}^{\infty }C_{n,2k}^{\left( cal\right) }\left( \epsilon ,\omega
\right) \text{ }x^{2k},  \label{mb}
\end{equation}%
where the expansion coefficients $C_{n,2k}^{\left( cal\right) }\left(
\epsilon ,\omega \right) $ depending on the coupling moduli,%
\begin{equation}
C_{n,2k}^{\left( cal\right) }\left( \epsilon ,\omega \right)
=\sum_{j=0}^{k}\left( \frac{(-\omega )^{k}C_{n}^{j}}{(k-j)!n!2^{k-j}}\frac{%
\Gamma (n+\epsilon +\frac{1}{2})}{\Gamma (j+\epsilon +\frac{1}{2})}\right) ,
\label{mc}
\end{equation}%
with $C_{n}^{j}=n!/j!\left( n-j\right) !$ and where for later use note that $%
C_{n,0}^{\left( cal\right) }\neq 0$. Though forbidden by the reality
condition of total wave $\Psi \left( x\right) $, the manifest discrete
symmetry $x\leftrightarrow \left( -x\right) $ of solution (\ref{mb}) is not
hazardous; it is a symmetry of the Calogero hamiltonian operator (\ref{ca}).
Under the exchange of the position $x_{1}\leftrightarrow x_{2}$, the complex
wave function maps as,%
\begin{equation}
\Psi \left( x_{1},x_{2}\right) =e^{i\pi \epsilon }\Psi \left(
x_{2},x_{1}\right) ,  \label{gro}
\end{equation}%
showing that, for $\epsilon $ different from even integer, the two particles
obey a non trivial statistics. (\textbf{b}) The second thing we want to note
is that the coefficients $C_{n,0}^{\left( cal\right) }$ and $C_{n,2}^{\left(
cal\right) }$of the series (\ref{mb}) are related as follows,%
\begin{equation}
C_{n,2}^{\left( cal\right) }=\frac{-E_{n}^{\left( cal\right) }}{\left(
2\epsilon +1\right) }C_{n,0}^{\left( cal\right) }.  \label{en}
\end{equation}%
From eq(\ref{mb}), one sees that above relation may be also rewritten as $%
E_{cal}=-\left( 2\epsilon +1\right) \frac{\Phi _{cal}^{\prime \prime }\left(
0\right) }{\Phi _{cal}\left( 0\right) }$. (\textbf{c}) The third thing we
want to comment is that quantity $\left( f^{\prime 2}-2\frac{\varepsilon }{x}%
f^{\prime }-f^{\prime \prime }-2U\right) $ appearing in eq(\ref{ee}) is
constant, say equal to $K$. Its solution is obviously given by $U=\frac{%
\omega ^{2}}{2}x^{2},$ $f=\frac{\omega }{2}x^{2}$ and $K=-\left( 2\epsilon
+1\right) $. Note also if we take $f=\left( \epsilon -1\right) \ln x$, one
gets precisely $U=\frac{g}{2x^{2}}$ which is nothing but the first
factorisation we have used to absorb Calogero term in the hamiltonian.

\section{Calogero model with a Yukawa like potential}

\qquad In this section, we consider the generalization of previous analysis
to the case where the Calogero particles are electrically charged and are
moreover strongly correlated. The interaction describing this particular
behavior is given by the following Yukawa like potential $\cite{16}$,%
\begin{equation}
V_{Yuk}\left( x\right) =\frac{2\alpha }{x}\exp \left( -\frac{x}{\lambda }%
\right) ,\qquad x\sqrt{2}=\left\vert x_{1}-x_{2}\right\vert ,
\end{equation}%
where $\alpha $ is a coupling constant and $\lambda $ is the Debye length.
Note that for the range $x<\lambda $, the two leading terms of $%
V_{Yuk}\left( x\right) $ are respectively given by the usual Coulombian term
$\frac{2\alpha }{x}$\ and a constant $\frac{-2\alpha }{\lambda }$. The
presence of the constant $\frac{-2\alpha }{\lambda }$ shows that energy
spectrum $E$ of this generalized system \ should read as $E=E_{cal}-\frac{%
2\alpha }{\lambda }+$ \textit{other contributions} coming from Coulomb term
and remaining polynomial fluctuations ($\exp \left( -\frac{x}{\lambda }%
\right) \sim 1-\frac{x}{\lambda }+\frac{1}{2}\left( \frac{x}{\lambda }%
\right) ^{2}$). For $x>\lambda $, the spectrum of the quantum system is
mainly given by the usual Calogero one ($\frac{2\alpha }{x}\exp \left( -%
\frac{x}{\lambda }\right) \sim 0$). The Shrodinger equation describing the
quantum dynamics of these interacting particles reads, in local coordinates $%
x_{1}$ and $x_{2}$, as,
\begin{equation}
\left( -\frac{1}{2}\sum_{i=1}^{2}\frac{\partial ^{2}}{\partial x_{i}^{2}}%
+V\left( x_{1},x_{2}\right) \right) \Psi \left( x_{1},x_{2}\right) =E\Psi
\left( x_{1},x_{2}\right) ,  \label{ka1}
\end{equation}%
where $\Psi $ is the total wave function, $V$ is the total translation
invariant potential given by,%
\begin{equation}
V\left( x\right) =\frac{\omega ^{2}}{2}x^{2}+\frac{g}{2x^{2}}+\frac{\sqrt{2}%
\alpha }{x}\exp \left( -\frac{\sqrt{2}x}{\lambda }\right) ,  \label{ka2}
\end{equation}%
and $E$ is the total energy eigenvalue depending on the potential moduli $%
\omega ,$ $g,$ $\alpha $ and $\lambda $. Note that, like for Calogero
system, transformation generated by space scale dilatation $x\rightarrow
\zeta x,$ maps the hamiltonian $\mathcal{H}$ as follows $\mathcal{H}%
\rightarrow \zeta ^{-2}\mathcal{H}$ provided that couplings are mapped as $%
\omega \rightarrow \zeta ^{-2}\omega ,$ $g\rightarrow g,$ $\alpha
\rightarrow \zeta ^{-1}\alpha $ and $\lambda \rightarrow \zeta \lambda $.
Note moreover that the limit $\alpha $ small is also equivalent to large
values of the relative distance $x$ with respect to the Debye length $%
\lambda $ ($x>>\lambda $). In this region, the additional Yukawa like
interaction $\frac{\sqrt{2}\alpha }{x}\exp \left( -\frac{\sqrt{2}}{\lambda }%
x\right) $ may be neglected.

\qquad To get the discrete spectrum $\left\{ E_{n},\Psi _{n}\left( x\right)
\right\} $ of the wave equation eqs(\ref{ka1},\ref{ka2}), we shall follow
the Calogero method outlined in previous section and the moment technique to
be introduced later on. Using \textit{classical} properties of the wave
function and translation invariance ($x_{i}\rightarrow x_{i}+cst$), we can
thus decompose $\Psi \left( x\right) $ in three factors as before $\Psi
\left( x\right) =x^{\varepsilon }\exp \left( -\frac{\omega }{2}x^{2}\right)
P\left( x\right) $ where a priori $\epsilon $ is positive; but quantum
effects requires that it should be as $\epsilon >-\frac{1}{2}$ as we show
later, see also eq(\ref{con}) and $\cite{17}$. In the first step, the
factorisation of $\Psi $ reads as follows $\Psi \left( x\right) =x^{\epsilon
}\mathcal{\Phi }(x)$. Like in Calogero analysis, this factorisation is also
used to kill the non linear term $\frac{g}{x^{2}}$ by making an appropriate
choice of $\epsilon $ parameter ($\epsilon =\epsilon \left( g\right) $).
Finding $\Psi \left( x\right) $\ is then equivalent to determining the $%
\epsilon $ parameter in terms of potential moduli and the unknown function $%
\mathcal{\Phi }(x)$. Substituting $\Psi ^{\prime \prime }=x^{\epsilon }%
\mathcal{\Phi }^{\prime \prime }+2\epsilon x^{\epsilon -1}\mathcal{\Phi }%
^{\prime }+\epsilon (\epsilon -1)x^{\epsilon -2}\mathcal{\Phi }$ in eq(\ref%
{ka1}), Schr\"{o}dinger equation becomes,%
\begin{equation}
\mathcal{\Phi }^{\prime \prime }+\frac{2\epsilon }{x}\mathcal{\Phi }^{\prime
}+\left[ \frac{\epsilon (\epsilon -1)-g}{x^{2}}-\omega ^{2}x^{2}-\frac{2%
\sqrt{2}\alpha }{x}\exp \left( \frac{-x\sqrt{2}}{\lambda }\right) +2E\right]
\mathcal{\Phi }=0.  \label{ka4}
\end{equation}%
This is a second order differential equation with non constant coefficients
which, as we know, its solution isn't a simple matter. However, eq(\ref{ka4}%
) may be simplified a little bit by working, like in Calogero model, in a
special region of the coupling constant moduli space;
\begin{equation}
\epsilon (\epsilon -1)-g=0\text{,\qquad }\alpha ,\lambda \text{ free moduli}.
\end{equation}%
Note that $g=g\left( \epsilon \right) $ is quadratic function in $\epsilon $
and so it has two zeros namely $\epsilon _{-}\left( g=0\right) =0$ and $%
\epsilon _{+}\left( g=0\right) =1$, see also figure, \bigskip
\begin{figure}[tbh]
\begin{center}
\epsfxsize=6cm \epsffile{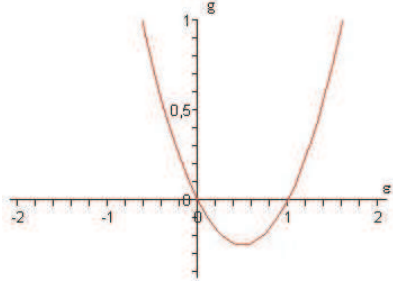}
\end{center}
\caption{{\protect\small \textit{This figure describes the variation of
Calogero coupling constant }}${\protect\small \mathit{g}}${\protect\small
\textit{\ in term of the statistical parameter }}$\protect\epsilon $%
{\protect\small \textit{. As one sees there is two regions for Calogero
coupling g, attractive and repulsive. In the interface }}${\protect\small
\mathit{g=0}}${\protect\small \textit{, we have two solutions }}%
{\protect\small \textit{\ $\left( \protect\epsilon _{-},g\right) =\left(
0,0\right) $\textit{\ } and $\left( \protect\epsilon _{+},g\right) =\left(
1,0\right) $,}} {\protect\small \textit{\ which, by \textit{e}q(\protect\ref%
{gro}), shows that they correspond to Bose and Fermi statistics respectively}%
}.{\protect\small \textit{\ For generic values of g, we have the two
branches }}$\protect\epsilon _{\pm }\left( g\right) =\frac{1\pm \protect%
\sqrt{1+4g}}{2}$.}
\label{figade}
\end{figure}
As $\epsilon $ is the basic parameter in this analysis, it is then
interesting to express it like $\epsilon =\epsilon \left( g\right) $; but
this leads however to two possible solutions $\epsilon _{\pm }\left(
g\right) $ given by,%
\begin{equation}
\epsilon _{-}\left( g\right) =\frac{1-\sqrt{1+4g}}{2},\qquad \epsilon
_{+}\left( g\right) =\frac{1+\sqrt{1+4g}}{2}.  \label{kes}
\end{equation}%
Generally speaking, one should distinguish between these two cases; however
this not necessary since these values are related as $\epsilon _{+}\left(
g\right) +\epsilon _{-}\left( g\right) =1$. One derives the quantum spectrum
for one of the cases, say $\epsilon =\epsilon _{+}\left( g\right) $, then
deduces the other one by using the identity $\epsilon _{-}=1-\epsilon _{+}$.
The second step is to factor $\Phi $ as $\Phi \left( x\right) =P\left(
x\right) \exp \left( -\frac{\omega }{2}x^{2}\right) $ in order to absorb the
harmonic term $\omega ^{2}x^{2}$. By substitution, the previous equation
becomes%
\begin{equation}
P^{\prime \prime }+\left( \frac{2\epsilon }{x}-2x\omega \right) P^{\prime
}+\left( 2E-\left( 2\epsilon +1\right) \omega -\frac{2\sqrt{2}\alpha }{x}e^{%
\frac{-x\sqrt{2}}{\lambda }}\right) P=0,  \label{erf}
\end{equation}%
where obviously $\epsilon $ stands for $\epsilon _{\pm }\left( g\right) =%
\frac{1\pm \sqrt{1+4g}}{2}$. Strictly speaking we have two wave equations to
solve and then two discrete spectrum to get; we shall refer to them as $%
\left\{ \Psi _{n}^{\pm },\text{ }E_{n}^{\pm }\right\} $. Note moreover that
the two above factorisation of the field variable $\Psi \left( x\right) $
implies, amongst others, the two following: (\textbf{i}) The first change of
variables fixes the parameter $\epsilon $ as $\epsilon _{\pm }=\frac{1\pm
\sqrt{1+4g}}{2}$ whose reality property requires that $g\geq -\frac{1}{4}$
showing that there are various phases, an attractive one associated to the
moduli space $-\frac{1}{4}\leq g<0$ and which may be interpreted as a
confining phase, two repulsive ones given by $g>0$ with $-\frac{1}{2}%
<\epsilon _{-}\leq 0$ and $\epsilon _{+}\geq 1$ and finally two interfaces
associated with the points $\left( \epsilon _{-},g\right) =\left( 0,0\right)
$ and $\left( \epsilon _{+},g\right) =\left( 1,0\right) $. Note that quantum
mechanically, the classical argument $\Psi \left( x\right) =0$ for $x=0$
leading to $\epsilon >0$ is not a quantum principle. The right condition is
to require rather that $\Psi \left( x\right) $ is a Hilbert space wave
function with a convergent integral $\int dx\left\vert \Psi \left( x\right)
\right\vert ^{2}<\infty $. Following $\cite{16}$, wave functions such as $%
\Psi \left( x\right) =x^{\epsilon }\Phi \left( x\right) ,$ with $\epsilon >%
\frac{-1}{2},$ are also Hilbert space solutions despite the apparent
divergence at $x=0$. This feature is a typical quantum mechanical property
which may be derived in several manners; one of which is the elaborated one
developed in $\cite{16}$ using canonical operator formalism; a second one is
given by the direct method to be developed below and leading to eq(\ref{27}%
). To have a quick idea on the rule behind this $\epsilon $ shift to left,
we present here below a naive way to see it. The argument relies on
considering eq(\ref{erf}) and requires vanishing of fundamental energy for
arbitrary values of relative position $x$. This is a typical feature
dictated by renormalisability of quantum systems and allow to redefine
energy origin by absorbing the constant term. Recall that in quantum
mechanics, the fundamental energy constant term captures quantum effects and
is one of the sources of divergences. In canonical quantization for
instance, this term has some thing to do with operators normal ordering and
is absorbed in the renormalisation process shifting energy as $E\geq 0$. For
our concern, it is enough to look at boundaries of the system ($\left\vert
x\right\vert \rightarrow \infty $) where Yukawa like interaction may be
ignored. The corresponding energy positivity condition $E\geq 0$ is then the
same as in Calogero case and reads as $\frac{E}{2\omega }\geq \frac{%
2\varepsilon +1}{4}$ or equivalently $2\epsilon +1>0$. In general, one can
prove that for the case of $N$ particles, this condition extends as $%
N\epsilon +1>0$ showing that we should have $\epsilon >\frac{-1}{N}$. In
what follow we shall give an other way to deal with this quantum shift. (%
\textbf{ii}) The second factorisation of the wave function tells us that as
far as moduli space of coupling constants is concerned, one may also
consider the moduli space region where the term $\frac{\sqrt{2}\alpha }{x}%
\exp \left( \frac{-x\sqrt{2}}{\lambda }\right) $ can be treated as a
perturbation around Calogero potential ($\frac{\alpha }{\lambda }<<1$).
There, the spectrum $\left\{ E_{n}\left( \epsilon ,\omega ,\alpha ,\lambda
\right) ,\mathcal{P}_{n}\left( x;\epsilon ,\omega ,\alpha ,\lambda \right)
\right\} $ we are looking for should reduce to Calogero one, plus
fluctuations as shown below;
\begin{eqnarray}
\mathcal{P}_{n}\left( x;\epsilon ,\omega ,\alpha ,\lambda \right) &=&%
\mathcal{P}_{n}^{\left( cal\right) }\left( x;\epsilon ,\omega \right) +%
\mathcal{O}\left( \alpha ,\frac{1}{\lambda }\right) ,  \notag \\
E_{n}\left( \epsilon ,\alpha ,\lambda \right) &=&E_{n}^{\left( cal\right)
}\left( \epsilon ,\omega \right) +\mathcal{O}\left( \alpha ,\frac{1}{\lambda
}\right) ,
\end{eqnarray}%
where $\mathcal{P}_{n}^{\left( cal\right) }(x;\epsilon ,\omega )$ and $%
E_{n}^{\left( cal\right) }$ are the exact Calogero spectrum eq(\ref{ma});
see also conclusion section. The normalisation condition of the total wave
function $\Psi $, which, by using the factorisation $\Psi \left( x\right)
=x^{\epsilon }P\left( x\right) \exp \left( -\frac{\omega }{2}x^{2}\right) $,
reads as,%
\begin{equation}
\int_{0}^{\infty }\left\vert \Psi \left( x\right) \right\vert
^{2}dx=\int_{0}^{\infty }x^{2\epsilon }P^{2}\left( x\right) e^{-\omega
x^{2}}dx,\qquad \epsilon >-\frac{1}{2},  \label{co}
\end{equation}%
requires convergence of the integral. For $x=\infty $, such condition is
ensured provided that $x^{2\epsilon }P^{2}\left( x\right) <e^{-\omega x^{2}}$
and for $x=0$, the quantity $\frac{x^{2\epsilon +1}}{2\epsilon +1}%
P^{2}\left( 0\right) $ should be finite. Both these conditions are fulfilled
for Laguerre polynomials associated with Calogero model. In the case where
Yukawa like coupling are implemented, the function $P\left( x\right) $ is no
longer polynomial but has expansion type $\sum_{s\geq 0}p_{s}x^{s}$
involving positive modes only; that is with a good behavior near $x=0$.
Under this assumption we can prove that eq(\ref{co}) converges as well for
the range $\epsilon >-\frac{1}{2}$ exactly as in Calogero system.\ Let us
give some details on the proof. First note there are two sources for
possible divergences of eq(\ref{co}), one at $x=+\infty $ and the other at $%
x=0$. At infinity, Yukawa like potential $\frac{1}{x}e^{-ax}$ is negligible
and the system is mainly the Calogero one where, as noted before, solutions
are known for the range $\epsilon >-\frac{1}{2}$. In this case $P\left(
x\right) $ is a polynomial eq(11) and its convergence is ensured by the
gaussian factor whatever the value of $\epsilon $ is. At $x=0$ however, the
situation is some how subtle; we have $e^{-\omega x^{2}}\sim 1-\omega
x^{2}+O\left( x^{4}\right) $ and so a possible divergence may come from low
modes of the expansion of $P\left( x\right) $ which may be approximated in
this region as $p_{0}+p_{1}x+O\left( x^{2}\right) $ independently on whether
$P\left( x\right) $ is a polynom or not; the only requirement is that $%
P\left( x\right) $ should have positive modes only. Therefore, we have $%
x^{2\epsilon }P^{2}\left( x\right) e^{-\omega x^{2}}\sim
p_{0}^{2}x^{2\epsilon }+2p_{0}p_{1}x^{2\epsilon +1}+O\left( x^{2\epsilon
+2}\right) $, which diverges for $\epsilon \leq \frac{-1}{2}$ since,
\begin{equation}
\int \left\vert \Psi \left( x\right) \right\vert ^{2}dx\simeq \frac{p_{0}^{2}%
}{2\epsilon +1}x^{2\epsilon +1}+\frac{2p_{0}p_{1}}{2\epsilon +2}x^{2\epsilon
+2}+\frac{p_{1}^{2}-p_{0}^{2}\omega }{2\epsilon +3}x^{2\epsilon +3}+O\left(
x^{2\epsilon +4}\right) .  \label{27}
\end{equation}%
At $2\epsilon +1=0$, one has a logarithmic divergence; it corresponds to a
critical point describing collapsing particles as interpreted in $\cite{16}$%
. For $2\epsilon +2=0,$ one has both linear and logarithmic divergences.
However in the range $\epsilon >\frac{-1}{2}$ the situation is healthy like
in Calogero model. Note that convergence of $\int \left\vert \Psi \left(
x\right) \right\vert ^{2}dx$ is also sensitive to the values of the modes $%
p_{s}$. For $p_{0}=0$ for instance, even the logarithmic divergence at $%
\epsilon =\frac{-1}{2}$ disappear and, as one sees on eq(\ref{27}), is
shifted to the moduli space point $\epsilon =\frac{-3}{2}$.

\begin{theorem}
:\qquad \newline
(\textbf{a}) The discrete energy eigenvalues $E_{n}$ of generalized Calogero
system with electrically charged particle and strongly correlated described,
in addition to Calogero term, by a Yukawa like interaction eq(\ref{ka4}) are
given by
\begin{equation}
E_{n}\left( \epsilon ,\alpha ,\lambda \right) =-\left( 2\epsilon +1\right)
\frac{\Phi _{n}^{\prime \prime }\left( 0\right) }{\Phi _{n}\left( 0\right) }%
+f\left( \alpha ,\lambda \right)
\end{equation}%
where $\Phi _{n}\left( 0\right) $ and $\Phi _{n}^{\prime \prime }\left(
0\right) $\ are the values of the wave function $\Phi \left( x\right) $ and
its second derivative at $x=0$ and where $f\left( \alpha ,\lambda \right) $
is equal to $\frac{2\alpha \left( \lambda \alpha -\epsilon \right) }{%
\epsilon \lambda }$.\newline
(\textbf{b}) The total wave functions $\Psi _{n}\left( x\right) =x^{\epsilon
}\Phi _{n}\left( x\right) $ of this generalized Calogero system is
completely determined up to the knowledge of the field (zero mode) values $%
\Phi _{n}\left( x=0\right) $ and $\Phi _{n}^{\prime \prime }\left(
x=0\right) $.
\end{theorem}

\qquad Having in mind the method of solving standard Calogero wave equation
presented in section 2 and convergence of (\ref{co}) for $P\left( x\right) $
with an expansion type $\sum_{s\geq 0}p_{s}x^{s}$ involving positive modes,
one may proceed ahead to establish this theorem by mimicking the same
method. The idea is to start from eq(\ref{ka4}) which we rewrite as,
\begin{eqnarray}
\mathcal{\Phi }_{n}^{\prime \prime }+\frac{2\epsilon }{x}\mathcal{\Phi }%
_{n}^{\prime }+\left[ 2E_{n}-\omega ^{2}x^{2}-\frac{2\sqrt{2}\alpha }{x}\exp
\left( \frac{-x\sqrt{2}}{\lambda }\right) \right] \mathcal{\Phi }_{n} &=&0,
\notag \\
P^{\prime \prime }\left( x\right) +\left( \frac{2\varepsilon }{x}-2\omega
x\right) P^{\prime }\left( x\right) +\left( 2E-\left( 2\epsilon +1\right)
\omega -U\right) P\left( x\right) &=&0,  \label{es}
\end{eqnarray}%
where $U=\frac{2\sqrt{2}\alpha }{x}e^{\frac{-x\sqrt{2}}{\lambda }}$ and try
to put it in a form type (\ref{la}-\ref{lb}).

To prove the theorem, we shall develop a method inspired from the solution (%
\ref{mb}) and an expansion of the wave function as a formal series. The key
idea may be summarized as follows: first use the feature $\int_{-\infty
}^{\infty }dx\left\vert \Psi _{n}\left( x\right) \right\vert ^{2}<\infty $
eq(\ref{co}) and the factorisation $\Psi _{n}\left( x\right) =x^{\epsilon
}P_{n}\exp \left( -\frac{\omega }{2}x^{2}\right) $ to build the wave
functions $\Psi _{n}$. Second benefit from known properties of Calogero wave
function, in particular its development as a series (\ref{mb},\ref{mc}) with
positive modes which we suppose also valid in presence of Yukawa like
interaction,%
\begin{equation}
\Phi _{n}\left( x\right) =\sum_{r\geq 0}C_{n,r}\left( \epsilon ,\omega
,\alpha ,\lambda \right) \text{ }x^{r},\qquad \epsilon >0,
\end{equation}%
where\ the $C_{n,k}$ coefficients, depending on the coupling moduli, are the
new unknown quantities which have to be determined. This means that, like $%
\Psi _{n}^{\left( cal\right) }\left( x\right) =x^{\epsilon }\Phi ^{\left(
cal\right) }\left( x\right) $, the wave $\Psi _{n}\left( x\right)
=x^{\epsilon }\Phi _{n}\left( x\right) $ may be expanded in a formal series
as,%
\begin{equation}
\Psi _{n}\left( x\right) =\sum_{r\geq 0}C_{n,r}\left( \epsilon ,\omega
,\alpha ,\lambda \right) \text{ }x^{r+\epsilon }.  \label{y}
\end{equation}%
To that purpose, note that in a distribution sense, the integral $%
\int_{-1}^{1}\frac{1}{m+1}x^{m}dx=\int_{-1}^{1}d\left( x^{m+1}\right) $ is
usually equal to zero except for the case $m=-1$ where the integral is
different from zero. Using this property, which may be thought of as a real
version of Cauchy theorem of complex analysis, we can inverse eq(\ref{y}) as
follows,%
\begin{equation}
C_{n,l}\left( \epsilon ,\omega ,\alpha ,\lambda \right)
=\int_{-1}^{1}x^{-l-1}\Phi _{n}\left( x\right) =\frac{1}{l!}\Phi ^{\left(
l\right) }\left( 0\right) ,\qquad l\geq 0.
\end{equation}%
To get the explicit expression of the new variables $C_{n,l}\left( \epsilon
,\omega ,\alpha ,\lambda \right) $, we start from the differential equation,%
\begin{equation}
x^{2}\mathcal{\Phi }_{n}^{\prime \prime }+2\epsilon \left( x\mathcal{\Phi }%
_{n}^{\prime }\right) +\left[ 2E_{n}x^{2}-\omega ^{2}x^{4}-2\sqrt{2}\alpha
x\exp \left( \frac{-x\sqrt{2}}{\lambda }\right) \right] \mathcal{\Phi }%
_{n}=0.  \label{k6}
\end{equation}%
Putting the expansion $\mathcal{\Phi }_{n}\left( x\right)
=\sum_{k=0}^{\infty }C_{n,k}$ $x^{k}$ back into eq(\ref{k6}) and using the
development $\exp \left( \frac{-x\sqrt{2}}{\lambda }\right) ,$ namely $%
\sum_{m=0}^{\infty }\frac{\left( -\right) ^{m}2^{\frac{m}{2}}}{\lambda ^{m}m!%
}x^{m}$, we get after term rearranging the following algebraic equation,%
\begin{equation}
\sum_{k\geq 0}A_{n,k}\text{ }x^{k}=0  \label{k7}
\end{equation}%
where the first four $A_{n,k}$\ coefficients are as $A_{n,0}=\left[ \epsilon
(\epsilon -1)-g\right] C_{n,0}$, $A_{n,1}=\left( 2\epsilon +\epsilon
(\epsilon -1)-g\right) C_{n,1}-2\sqrt{2}\alpha C_{n,0}$ together with
\begin{eqnarray}
A_{n,2} &=&\left( 2\left( 2\epsilon +1\right) +\epsilon (\epsilon
-1)-g\right) C_{n,2}-2\left[ \sqrt{2}\alpha C_{n,1}-\left( E_{n}+\frac{%
2\alpha }{\lambda }\right) C_{n,0}\right] ,  \label{32} \\
A_{n,3} &=&\left( 3\left( 2\epsilon +2\right) +\epsilon (\epsilon
-1)-g\right) C_{n,3}-2\left[ \sqrt{2}\alpha C_{n,2}-\left( E_{n}+\frac{%
2\alpha }{\lambda }\right) C_{n,1}+\frac{\sqrt{2}\alpha }{\lambda ^{2}}%
C_{n,0}\right] .  \notag
\end{eqnarray}%
and the remaining others ($k\geq 4$) read as $A_{n,k}=\left[ k\left(
2\epsilon +k-1\right) +\epsilon (\epsilon -1)-g\right] C_{n,k}-D_{n,k}$,
with $D_{n,k}$s given by,%
\begin{equation}
D_{n,k}=2\alpha \sum_{m=0}^{k-1}\frac{\left( -\right) ^{m}2^{\frac{m+1}{2}}}{%
m!\lambda ^{m}}C_{n,k-m-1}+\omega ^{2}C_{n,k-4}-2E_{n}C_{n,k-2}
\end{equation}%
The solution of eq(\ref{k7}) for arbitrary $x$ is obtained by requiring the
vanishing of the $A_{n,k}$ coefficients which give in turn the following
recurrent constraint eqs on the $C_{n,k}$\ functions,%
\begin{eqnarray}
0 &=&\left[ \epsilon (\epsilon -1)-g\right] C_{n,0},  \notag \\
C_{n,1} &=&\frac{2\alpha \sqrt{2}}{2\epsilon +\epsilon (\epsilon -1)-g}%
C_{n,0},  \notag \\
C_{n,2} &=&\frac{2\left[ \sqrt{2}\alpha C_{n,1}-\left( E_{n}+\frac{2\alpha }{%
\lambda }\right) C_{n,0}\right] }{2\left( 2\epsilon +1\right) +\epsilon
(\epsilon -1)-g},  \label{k10} \\
C_{n,3} &=&\frac{2\sqrt{2}\alpha C_{n,2}-2\left( E_{n}+\frac{2\alpha }{%
\lambda }\right) C_{n,1}+2\frac{\sqrt{2}\alpha }{\lambda ^{2}}C_{n,0}}{%
3\left( \epsilon +1\right) +\epsilon (\epsilon -1)-g},  \notag
\end{eqnarray}%
and for $k\geq 4,$
\begin{equation}
C_{n,k}=\frac{2\alpha \sum_{m=0}^{k-1}\frac{\left( -\right) ^{m}2^{\frac{m+1%
}{2}}}{m!\lambda ^{m}}C_{n,k-m-1}+\omega ^{2}C_{n,k-4}-2E_{n}C_{n,k-2}}{%
k(2\epsilon +k-1)+\epsilon (\epsilon -1)-g}  \label{k11}
\end{equation}%
From these recurrent relations, one sees that in the case $\epsilon \neq 0$
the elements of the infinite set $\left\{ C_{n,k}\right\} $ are all of them
function of $C_{n,0}$, the zero mode of the expansion (\ref{y}), and the
energy $E_{n}$. Notice that in the case $\epsilon =0$, one is left with
Yukawa like model since classically there is no Calogero interaction. In
this situation and using the relations $A_{n,0}=\left[ \epsilon (\epsilon
-1)-g\right] C_{n,0}$, $A_{n,1}=\left( 2\epsilon +\epsilon (\epsilon
-1)-g\right) C_{n,1}-2\sqrt{2}\alpha C_{n,0}$ together with $A_{n,2}$ and $%
A_{n,3}$ given by eqs(\ref{32}), we have either $C_{n,0}\neq 0$; but $\alpha
=0$ or $C_{n,0}=0$ and $\alpha \neq 0$. In the second case, one should have $%
C_{n,1}\neq 0$ otherwise the wave function vanishes. Non trivial solution%
\footnote{%
The moduli space poles at $2\epsilon =\left( 1-k\right) $, $k=2,3,...,$
should be compared with those of eq(\ref{27}).} requires $C_{n,0}\neq 0$;
the first equation of (\ref{k10}) fixes then the Calogero $g$ coupling
constant as in the case of absence of Yukawa like coupling,
\begin{equation}
\epsilon (\epsilon -1)=g.  \label{k12}
\end{equation}%
Putting this relation back into above eqs(\ref{k10},\ref{k11}), we get the
following reduced constraint eqs,%
\begin{eqnarray}
C_{n,1} &=&\frac{\alpha \sqrt{2}}{\epsilon }C_{n,0},  \notag \\
C_{n,2} &=&\frac{C_{n,0}}{2\epsilon +1}\left( \frac{2\alpha ^{2}}{\epsilon }-%
\frac{2\alpha }{\lambda }-E\right) ,  \label{k13} \\
C_{n,3} &=&\frac{\sqrt{2}\alpha C_{n,0}}{3\left( \epsilon +1\right) }\left[
\frac{2\alpha ^{2}}{\epsilon \left( 2\epsilon +1\right) }-\frac{\left(
3\epsilon +1\right) \left( E_{n}+\frac{2\alpha }{\lambda }\right) }{\epsilon
\left( 2\epsilon +1\right) }+\frac{1}{\lambda ^{2}}\right] ,  \notag
\end{eqnarray}%
together with the following recurrent relations,%
\begin{equation}
C_{n,k}=\frac{2\alpha \sum_{m=0}^{k-1}\frac{\left( -\right) ^{m}2^{\frac{m+1%
}{2}}}{m!\lambda ^{m}}C_{n,k-m-1}+\omega ^{2}C_{n,k-4}-2E_{n}C_{n,k-2}}{%
k(2\epsilon +k-1)}.  \label{k14}
\end{equation}%
From eqs(\ref{k13}), one sees that for $\epsilon \neq 0$ one can express the
energy $E_{n}$ in terms of the coefficients $C_{n,0}$ and $C_{n,2}$ as shown
below,%
\begin{equation}
E_{n}=\left( \frac{2\alpha ^{2}}{\epsilon }-\frac{2\alpha }{\lambda }\right)
-\left( 2\epsilon +1\right) \frac{C_{n,2}}{C_{n,0}}.
\end{equation}%
The knowledge of $C_{n,0}$ and $C_{n,2}$ fixes then completely the spectrum $%
\left\{ \Psi _{n}\left( x\right) ,E_{n}\right\} $; but we have no exact way
to determine $C_{n,0}$ and $C_{n,2}$; except the use of approximation
methods. This is the case where the correlation between particles are strong
enough ($x<<\lambda $) so that the Yukawa like coupling may be interpreted
as a perturbation with respect to Calogero interaction. In this limit, the
ratio $\frac{C_{n,2}}{C_{n,0}}$ may be approximated by eq(\ref{en}) and
consequently the approximated energy reads as $E_{n}=\left( \frac{2\alpha
^{2}}{\epsilon }-\frac{2\alpha }{\lambda }\right) +E_{n}^{cal}$ with $%
\epsilon \neq 0$. As one sees $\Delta E_{n}=\Delta E_{n}^{cal}\sim \Delta
E_{n}^{osc}$ whatever positive integer $n$ is; it would be interesting to
check this behaviour by other methods. Progress in this matter will be
exposed in a future occasion. Note finally that the case $\epsilon =0$ with
non trivial wave function requires $C_{n,0}=0$, but $C_{n,1}\neq 0$ and the
previous energy relation get replaced by $2\alpha ^{2}-\frac{2\alpha }{%
\lambda }-3\frac{C_{n,3}}{C_{n,1}}$.

\section{Conclusion and discussions}

\qquad In this paper, we have studied an extension of one dimensional
Calogero model involving electrically charged particles with strong
correlations. These couplings are described, in addition to Calogero
potential $\frac{g}{2x^{2}}$, by a Yukawa like interaction $V_{Yuk}\left(
x\right) =\frac{2\alpha }{x}\exp \left( -\frac{x}{\lambda }\right) ,$ $x%
\sqrt{2}=\left\vert x_{1}-x_{2}\right\vert $. Though this system is not
completely integrable; we have shown that its discrete spectrum $\left\{
\Psi _{n}\left( x\right) =x^{\epsilon }\Phi _{n}\left( x\right)
,E_{n}\right\} $, $\epsilon >\frac{-1}{2},$ may be totally determined up to
fixing the value of the wave factor $\Phi _{n}\left( x\right) $ and $\Phi
_{n}^{\prime \prime }\left( x\right) $ at the singularity $x=0$ of the
potential; that is the zero mode $C_{n,0}$ and the second one $C_{n,2}$ of
the wave function expansion eq(\ref{y}). The quantum condition $\epsilon >%
\frac{-1}{2}$ is shown to be intimately related with the non zero value $%
\Phi _{n}\left( 0\right) $; see eq(\ref{27}). In the case of Calogero model,
the modes $C_{n,0}^{cal}$, $C_{n,2}^{cal}$ and energy $E_{n}^{cal}$ are all
of them fixed by a remarkable property of confluent hypergeometric functions
which are solution of eq(\ref{lb}) and which, for discrete spectrum, reduces
to Laguerre polynomials. In this study, we do not have a similar property
due to the presence of highly non linear potential interactions ($\exp
\left( -\frac{x}{\lambda }\right) $). We have focused on the case of two
identical particles $\left\{ x_{1},x_{2}\right\} $; but this is just a
matter of simplicity. Our analysis extends also to the case of a generic
number $N$ of identical particles $\left\{ x_{1},...,x_{N}\right\} $. There,
the wave function $\Psi $ depends on\ the $\frac{N\left( N-1\right) }{2}$
relative positions $x_{ij}=x_{i}-x_{j}$ ($\Psi \left( x\right) =\left( \Pi
_{i,j}x_{ij}\right) ^{\epsilon }\Phi \left( x_{12},...,x_{N-1,N}\right) $)
and the expansion of the factor $\Phi \left( x_{12},...,x_{N-1,N}\right) $
follows the same lines as for the case of two particles considered here.

\qquad In the end, we would like to add two comments. First note that for $%
g=\epsilon \left( \epsilon -1\right) $, one should distinguish the moduli
space regions: (i) $-\frac{1}{2}<\epsilon \leq 0$, (ii) $0\leq \epsilon \leq
1$ and (iii) $\epsilon >1$. Moreover solving $g=\epsilon \left( \epsilon
-1\right) $ as $\epsilon _{\pm }\left( g\right) $ eqs(\ref{kes}), one sees
that the corresponding spectrum $\left\{ \Psi _{n}^{\pm },E_{n}^{\pm
}\right\} $ may be split into $\left\{ \Psi _{2k+1}^{+},E_{2k+1}^{+}\right\}
$ and $\left\{ \Psi _{2k}^{-},E_{2k}^{-}\right\} $ with $k$ positive
integer. This splitting has a particularly interesting interpretation in the
moduli space region $g$ and $\alpha $ go to zero. Physically speaking, this
corresponds to large $x$ where the two potential components $\frac{g}{x^{2}}$
and $\frac{\alpha }{\left\vert x\right\vert }e^{-\frac{\left\vert
x\right\vert }{\lambda }}$ are negligible with respect to harmonic
interaction $\frac{\omega ^{2}}{2}x^{2}$. In this limit, the discrete
spectrum $\left\{ \Psi _{n}^{\pm },E_{n}^{\pm }\right\} $ should then tend
to the usual spectrum $\left\{ \Psi _{n}^{\left( osc\right) },E_{n}^{\left(
osc\right) }\right\} $ of one dimensional harmonic oscillator. Setting $%
g=\alpha =0$ in eq(\ref{ma}), one gets $E_{n}\left( \epsilon =0,\alpha
=0\right) =E_{n}^{cal}\left( \epsilon =0\right) =\left( 2n+\frac{1}{2}%
\right) \omega ,$ $n\in \mathbb{N}$; which at first sight seems not
reproducing the full energy spectrum of the harmonic oscillator that we
recall here $E_{n}^{\left( osc\right) }=\left( n+\frac{1}{2}\right) \omega ,$
$n\in \mathbb{N}$. The naive relation between $E_{n}^{cal}\left( \epsilon
=0\right) $ and $E_{n}^{\left( osc\right) }$ reads as,
\begin{equation}
E_{n}^{cal}\left( \epsilon =0\right) =E_{2n}^{\left( osc\right) },
\end{equation}%
and shows that apparently there are missing terms associated with the odd
integer modes $E_{2n+1}^{\left( osc\right) }$. In fact there is no
contradiction here since we have not implemented the full picture for the
limit $g=0$. In this case, one should distinguish two sectors $\epsilon
_{+}=1$, $g=0$ and $\epsilon _{-}=0$, $g=0$ whose respective energies $%
E_{n}^{\left( +,cal\right) }\left( g=0\right) $ and $E_{n}^{\left(
-,cal\right) }\left( g=0\right) $\ are given by $\lim_{g\rightarrow 0}\left[
\left( 2n+1\right) +\frac{1}{2}\sqrt{1+4g}\right] \omega =\left[ \left(
2n+1\right) +\frac{1}{2}\right] \omega $ and $\lim_{g\rightarrow 0}\left[
\left( 2n+1\right) -\frac{1}{2}\sqrt{1+4g}\right] \omega =\left[ 2n+\frac{1}{%
2}\right] \omega $. They read in terms of $E_{n}^{\left( osc\right) }$ as
follows,%
\begin{equation}
E_{n}^{\left( +,cal\right) }\left( g=0\right) =E_{2n+1}^{\left( osc\right)
},\qquad E_{n}^{\left( -,cal\right) }\left( g=0\right) =E_{2n}^{\left(
osc\right) }.
\end{equation}%
As one sees Calogero interaction splits harmonic oscillator energies into
two blocks $E_{2n+1}^{\left( osc\right) }$ and $E_{2n}^{\left( osc\right) }$
with a global separation $\left\vert E_{2n+1}^{\left( osc\right)
}-E_{2n}^{\left( osc\right) }\right\vert =\omega $ whatever the integer $n$
is. This is a remarkable feature which should deserve more investigation.
The second comment concerns the low mode structure of wave functions $\Psi
_{n}^{\pm }$ obtained from eqs(\ref{y}, \ref{k10}-\ref{k11}) by
implementation of eqs(\ref{kes}). In the limit $\epsilon _{+1}=1,$ $g=0$ and
the case $C_{n,0}^{+}\neq 0$, \ we have for leading terms $%
C_{n,1}^{+}=\alpha \sqrt{2}C_{n,0}^{+}$, $C_{n,2}^{+}=\frac{C_{n,0}^{+}}{3}%
(2\alpha ^{2}-\frac{2\alpha }{\lambda }-E_{n}^{+})$, $C_{n,3}^{+}=\frac{%
C_{n,0}^{+}}{6}(\frac{\sqrt{2}\alpha }{3}(2\alpha ^{2}-\frac{2\alpha }{%
\lambda }-E_{n}^{+})-\sqrt{2}\alpha (E_{n}^{+}+\frac{2\alpha }{\lambda })+%
\frac{\sqrt{2}\alpha }{\lambda ^{2}})$ together with the following modes $%
C_{n,k}^{+}=\frac{1}{k(k+1)}(2\alpha \sum_{m=0}^{k-1}\frac{(-)^{m}2^{\frac{%
m+1}{2}}}{m!\lambda ^{m}}C_{n,k-m-1}^{+}+\omega
^{2}C_{n,k-4}^{+}-2E_{n}^{+}C_{n,k-2}^{+}$) for $k\geq 4$. In the case $%
\epsilon _{-}=0,$ $g=0$; but $C_{n,0}^{-}=0$ and $C_{n,1}^{-}\neq 0$, we
have $C_{n,2}^{-}=\alpha \sqrt{2}C_{n,1}^{-},$ $C_{n,3}^{-}=\frac{C_{n,1}^{-}%
}{3}(2\alpha ^{2}-\frac{2\alpha }{\lambda }-E_{n}^{-})$, $C_{n,4}^{-}=\frac{%
C_{n,1}^{-}}{6}(\frac{\sqrt{2}\alpha }{3}(2\alpha ^{2}-\frac{2\alpha }{%
\lambda }-E_{n}^{-})-\sqrt{2}\alpha (E_{n}^{-}+\frac{2\alpha }{\lambda })+%
\frac{\sqrt{2}\alpha }{\lambda ^{2}})$ and for $k\geq 4$, $C_{n,k}^{-}=\frac{%
1}{k(k-1)}(2\alpha \sum_{m=0}^{k-1}\frac{(-)^{m}2^{\frac{m+1}{2}}}{m!\lambda
^{m}}C_{n,k-m-1}^{-}+\omega ^{2}C_{n,k-4}^{-}-2E_{n}^{-}C_{n,k-2}^{-}$).
Note that the modes in the two cases are related as $C_{n,k}^{+}(\epsilon
_{+}=1)C_{n,1}^{-}(\epsilon _{-}=0)=C_{n,k+1}^{-}(\epsilon
_{-}=0)C_{n,0}^{+}(\epsilon _{+}=1)$. At last note that near $x=0$, the wave
function $\Psi _{n}\left( x\right) =x^{\epsilon }\Phi _{n}\left( x\right) $
behaves as $x^{\epsilon }\left( C_{n,0}+C_{n,1}x+O\left( x^{2}\right)
\right) $. In the case where the leading modes are as $C_{n,0}=0$ and $%
C_{n,1}\neq 0$, all happens as if $\epsilon $ is shifted as $\widetilde{%
\epsilon }=\epsilon +1$ and $\widetilde{C}_{n,0}=C_{n,1}$.

\begin{acknowledgement}
\qquad The authors would like to thank prof A Intissar and A El Rhalami for
discussions. This research work is supported by the program Protars III
D12/25, CNRST.
\end{acknowledgement}

\end{document}